# Hadith Web Browser Verification Extension

Maged M. Eljazzar[1, a)], Mostafa Abdulhamid[2, b)], Mahmoud Mouneer[3, c)] and Ayman Salama[4, d)]

[1]Faculty of Engineering, Cairo University, Cairo, Egypt
[2]Cake Solutions Ltd, London, United Kingdom
[3] Faculty of petroleum and mining engineering, Suez University, Suez, Egypt
[4]Computer Science Division, Crops for the Future, University of Nottingham Malaysia Campus, Selangor, Malaysia

[a)] mmjazzar@ieee.org
[b)] lambda.mostafa@gmail.com
[c)] m.m.mouneer@gmail.com
[d)] ayman.s.mohamed@ieee.org

**Abstract.** Internet users are more likely to ignore Internet content verification and more likely to share the content. When it comes to Islamic content, it is crucial to share and spread fake or inaccurate content. Even if the verification process of Islamic content is becoming easier every day, the Internet users generally ignore the verification step and jump into sharing the content. "How many clicks away from users' results? ", this is the common question that is considered as a rule in modern website design. Internet users prefer the results to come to their page rather than to navigate it on their own. This paper presents a simple method of bringing hadith verification to the user web browser using web browser plugin.

## INTRODUCTION

Islamic knowledge transformation from books to structured Information technology is imminent. Number of Muslims that are obtaining their Islamic knowledge and Fatwa from Internet is significantly increasing. Advanced Machine learning [1] and Text mining [2] technologies should be widely used to accommodate the digital transformation of Islamic knowledge. Studying Social Network users' behaviour empowers the Islamic research tool to understand the Islamic data flow and to identify Internet threats on Islamic society.

The Pew Research Center published many survey results about Muslim's data [5-7]. The aim of its research is to review the opinions for a wide range of Islamic topics, from Jurisprudence (Fiqh), Western culture Impact on Muslim and the role of women in Islam. Some of these studies included in several Muslim and non-Muslim countries.

The number of Muslims Internet users in 39 countries across the Middle East, Europe, Asia and Africa forms a median of 18% of the total Internet users. According to Internet world stats [8] 246,700,900 population estimate for the Middle East in 2016. 141,489,765 Internet users for June/2016, 57.4% of the population. spacer 76,000,000

Facebook subscribers on Jun/2016, 30.8% penetration rate. In the Middle East, there are an estimated 141,489,765 Internet users, and the region has shown an impressive Internet usage growth rate of 4,207.4 %.between 2000 and 2016. The Internet usage growth rate of Muslim people requires a similar growth rate in the scientific research for the Islamic online content and its management.

Due to the sensitivity of Religious data, Sharia scientist spent several years studying and reading thousands of books to verify data accuracy. The main pressing problem is the percentage of sharing inaccurate data in social network. It could be in several forms, false illustration, wrong translation, or inaccurate data.

This work studies Hadith authentication in the Internet content especially in social network. This research is trying to identify a simple and quick method for the average Internet user to verify and authenticate various hadith content that the user stumbles in during surfing the Internet. Applying this method on a large scale will enable crowdsource analysis and big data mining for online Hadith content along with its users behavior.

## Literature review

Hadith is the major Islamic constitution source for Islam after Holy Quran. Recently computer related research in different Islamic aspects increased significantly with regards to the evolution of Computer Science. Extracting Islamic knowledge from traditional Islamic books using Data Mining techniques is widely used in Hadith classification, Isnad Studies, Knowledge Extraction from Hadith, and Takhreej Al Hadith Implementation [11-12]. Hadith Isnad Ontology Models is raising methodology that is recently used for testing the process of authenticating/judging Isnad [13]. [14] Moath present an innovative system for Hadith Isnad processing based on an Artificial Intelligence approach called Associative Classification (AC). Kawther et al. [15] use supervised learning to classify Hadith. They argue that the method of knowledge-extraction from Hadith is determined depending on the object of the Knowledge.

Some Islamic websites introduce Hadith applications to provides many tools for analyzing and indexing Hadiths. Those applications are also used as a reference to scientific Islamic researches. "Rewaia" encyclopedia [16] introduces 520,000 Hadith and 400 Volumes of Hadith related books. "Tarajim" encyclopedia contains the biographies of more than 150,000 narrators of Hadiths. These biographies include information about the Hadith narrator, date and place of birth, "Konia", date and place of death, teachers, students, etc., [17]. Shamela library is a significant resourceful application for both Hadith and different Islamic studies [18]. In hadith section, it includes information about narrators, Hadith Ma'tn and Isnad. Shamela Library lacks the automatic Hadith classification. It can not differentiate between Sahih and Da'ief automatically.

The automatic hadith authentication is becoming crucial with regards to the enormous use of Internet especially the social media. Dorar [19] introduce API (Application Program Interface) for developers to access Dorar database using JSON files via PHP or JavaScript. At last, Muhadith [20] is a cloud expert system that is designed to enable a computer to behave as a Hadith expert to differentiate the authentic hadith from unauthentic ones.

## Methodology

Hadith consists of two sections Mat'n and Isnad. Mat'n refers to the text of the Hadith, while Isnad means the chain of narrators to that Hadith. Hadith scientists divided the Traditions into categories according to the degree of authenticity and reliability, each category had to meet certain criteria. The categories are as follows:

1. Sahih: The most authentic Hadith.
2. Moothaq: Almost like the Sahih but the narration is not as strong as those of the Sahih.
3. Hassan: The fair Traditions although inferior in matter of authenticity.
4. Dha'eef: The weak ones which are not so reliable.

This research studies and implements hadith database designed based on Tradition Hadith books which are: Sahih Al-Bukhari, Sahih Muslim, Sunan Abu Dawod, Sunan Al-Termidhi, Sunan Ibn Majah,Sunan. These volumes of books contain more than five thousand Hadiths includes illustrations and classifications. Hadith Classifications and keywords associated to hadith are two main factors in the design of this database. The algorithm fetches the user search keywords for hadith and introduces the hadith illustration (if available) from the most authentic Hadith books. The algorithm can investigate the Hadith classification, and its trends according to these books. Using this method will enable the internet users to easily identify various Hadith classification that are found in any internet webpage. The algorithm can detect user behavior and needs and provide scientific based awareness to the Muslim community.

## MODEL

This research aims to collect Hadith search results from millions of Muslims internet users, especially from social networks. The system architecture was designed to minimize the processing time and to achieve the highest possible accuracy. This research conducts the development of a web browser plugin using Javascript. The plugin will automatically trigger remote database queries once the web browser user highlights possible Hadith text and initiate the plugin for Hadith classification checking as shown in fig.1. The figure shows the current upload of the plugin on Google Chrome plugins.

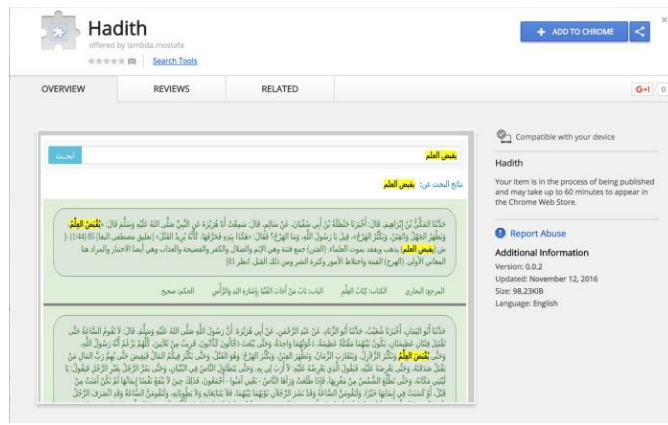

*Figure 1: How the user can search*

The user request will trigger the remote server to initiate the hadith classification algorithm. The algorithm will process the user highlighted text through the database. The result of the database query will be sent back to the user web browser's plugin. The plugin will provide the user with Hadith classification results through simple user interface. The process is illustrated in the below figure.2.

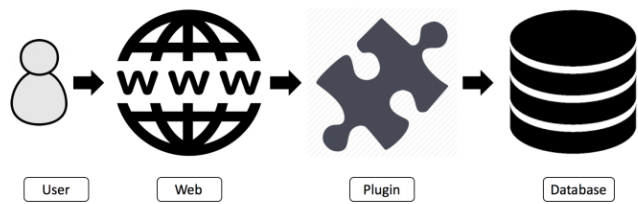

*Figure 2: Result UI*

This research is using Heroku Cloud Service to host the database and its application. The application is using Elasticsearch search engine to enable full text search within HTTP web interface. The main advantage of Elasticsearch is the multitenant-capable full-text search engine with an HTTP web interface and JSON documents Sitepoint[20]. This research uses mainly JavaScript for and JSON for communication between web browser plugin and the remote database in Heroku.

Every user request to the plugin is stored in the database along with its results for further analysis. This will allow the detection of the most sharable Hadiths on the internet along with its classification. This research doesn't aim to create new classification. It is using existing classification method based on authenticate Hadith books such as (Sahih Al-Bukhari, Sahih Muslim, etc.)., Each Hadith classification results can be different based on the used book. This problem and its possible solution are addressed in the future work of this research.

## EXPERIMENTAL RESULT

This research conduct an intensive collection of random Hadith test that are found on the internet. The test results show a good practice for the plugin. The plugin results were accurate based on the Hadith sources stated above. The result of the testing is shown in fig.3.

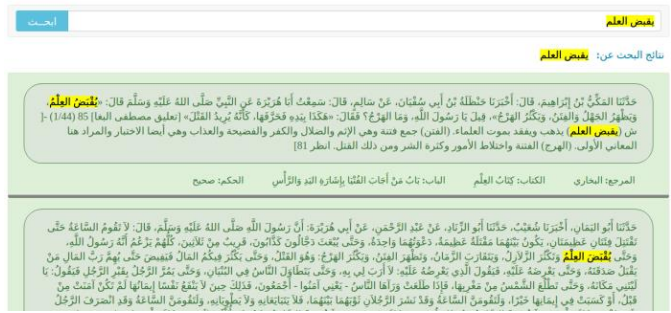

*Figure 3: Trend Results*

## Conclusion and Future work.

This research introduces a web based plugin to help internet users to verify various Hadith keywords found online using Hadith classification extraction from Traditional Hadith books (Sahih Al-Bukhari, Sahih Muslim, Sunan Abu Dawod, Sunan Al-Termidhi, Sunan Ibn Majah,Sunan)..The system store the user queries and its results for future analysis. This could be useful for Islamic studies to increase awareness in the trend of issues of Islamic communities. The study of this data will enable the researchers to identify the source of Dha'eef Hadith. The future work will include adding more books to increase the number of Hadith in the database. The current plugin takes the user highlighted text as an input. Commonly the internet users are using images to share possible Hadith keywords. The future work will include image processing to extract text from images that contains possible Hadith keywords. The expansion of this research work will include the development of an Application Program Interface (API) for the database to assist developers in Islamic communities to use the database.

Hadith classifications results can be different based on the used book. This problem and its possible solution are addressed in the future work of this research. One of the possible solution is the use Natural language processing

(NLP). Using NLP will enable the extraction of the main keywords of the results. In turn, it can enhance the classification.

There are several possible outcomes in the future work:
A. Investigating the users stored queries and its results will reveal the trend issues of Islamic society. It can identify which issue requires more attention from Islamic Communities?
B. Detecting the sources of Dha'eef Hadith
The stored results of all users' inquiries contain the source webpage of the Hadith. This will allow the detection of the possible sources of Dha'eef Hadith. This issue is crucial to online Muslim community to avoid possible planned attack on Islam using fake Hadith.
D. Ranking Hadith sources on the internet.

As the plugin based on the user request, The algorithm focuses to store the percentage of Sahih Hadith to Dha'eef Hadith in every website. The algorithm will rank the websites in general for its reputation for Hadith sharing types. The percentage is based on the number of Hadith search results that are in the Traditional books (Sahih Al-Bukhari, Sahih Muslim, etc.).

## ACKNOWLEDGMENTS

Shamela Library is resourceful tool; without it the research would have been more challenging.